\documentclass[11pt, preprint,longabstract]{aastex}
\slugcomment{Draft Version}

\begin{document}


\title{A STRONG RADIO BRIGHTENING AT THE JET BASE OF M~87 DURING THE ELEVATED
VERY-HIGH-ENERGY GAMMA-RAY STATE IN 2012}

\author{
K.~Hada\altaffilmark{1,2}, 
M.~Giroletti\altaffilmark{1}, 
M.~Kino\altaffilmark{3,4}, 
G.~Giovannini\altaffilmark{1,5}, 
F.~D'Ammando\altaffilmark{1},
C.~C.~Cheung\altaffilmark{6},
M.~Beilicke\altaffilmark{7}, 
H.~Nagai\altaffilmark{8}, 
A.~Doi\altaffilmark{4},
K.~Akiyama\altaffilmark{2,9}, 
M.~Honma\altaffilmark{2,10},
K.~Niinuma\altaffilmark{10}, 
C.~Casadio\altaffilmark{12},
M.~Orienti\altaffilmark{1}, 
H.~Krawczynski\altaffilmark{7}, 
J.~L.~G\'omez\altaffilmark{12}, 
S.~Sawada-Satoh\altaffilmark{2}, 
S.~Koyama\altaffilmark{2,4,9}, 
A.~Cesarini\altaffilmark{13}, 
S.~Nakahara\altaffilmark{14}, 
M.~A.~Gurwell\altaffilmark{15}
}

\affil{$^1$INAF Istituto di Radioastronomia, via Gobetti 101, I-40129 Bologna,
Italy; {\rm hada@ira.inaf.it}}

\affil{$^2$Mizusawa VLBI Observatory, National Astronomical Observatory of Japan,
Osawa, Mitaka, Tokyo 181-8588, Japan}

\affil{$^3$Korea Astronomy and Space Science Institute (KASI), 776 Daedeokdae-ro,
Yuseong-gu, Daejeon 305-348, Republic of Korea}

\affil{$^4$Institute of Space and Astronautical Science, Japan Aerospace
Exploration Agency, 3-1-1 Yoshinodai, Chuo, Sagamihara 252-5210, Japan}

\affil{$^5$Dipartimento di Fisica e Astronomia, Universit\`a di Bologna, via
Ranzani 1, I-40127 Bologna, Italy}

\affil{$^6$Space Science Division, Naval Research Laboratory, Washington, DC
20375, USA}

\affil{$^7$Physics Department and McDonnell Center for the Space Sciences,
Washington University, St. Louis, MO 63130, USA}

\affil{$^8$National Astronomical Observatory of Japan, Osawa, Mitaka, Tokyo
181-8588, Japan}

\affil{$^9$Department of Astronomy, Graduate School of Science, The University of
Tokyo, 7-3-1 Hongo, Bunkyo-ku, Tokyo 113-0033, Japan}

\affil{$^{10}$Department of Astronomical Science, The Graduate University for
Advanced Studies (SOKENDAI), 2-21-1 Osawa, Mitaka, Tokyo 181-8588, Japan}

\affil{$^{11}$Graduate School of Science and Engineering, Yamaguchi University, 1677-1
Yoshida, Yamaguchi, 753-8512, Japan} 

\affil{$^{12}$Instituto de Astrofisica de Andalucia, CSIC, Apartado 3004, 18080
Granada, Spain}

\affil{$^{13}$Department of Physics, University of Trento, I38050, Povo, Trento,
Italy}

\affil{$^{14}$Faculty of Science, Kagoshima University, 1-21-35 Korimoto, Kagoshima,
Kagoshima 890-0065, Japan}

\affil{$^{15}$Harvard-Smithsonian Center for Astrophysics, Cambridge MA 02138 USA}

%%%%%% BODY %%%%%%
\begin{abstract}
We report our intensive, high-angular-resolution radio monitoring observations of
the jet in M~87 with the VLBI Exploration of Radio Astrometry (VERA) and the
European VLBI Network (EVN) from February 2011 to October 2012, together with
contemporaneous high-energy (HE; 100~MeV$<E<$100~GeV) $\gamma$-ray light curves
obtained by the $\textit{Fermi}$ Large Area Telescope (LAT). During this period
(specifically from February 2012 to March 2012), an elevated level of the M~87
flux is reported at very-high-energy (VHE; $E>100$~GeV) $\gamma$-rays by
VERITAS. We detected a remarkable (up to $\sim$70\%) increase of the radio flux
density from the unresolved jet base (radio core) with VERA at 22 and 43~GHz
coincident with the VHE activity. Meanwhile, we confirmed with EVN at 5~GHz that
the peculiar knot HST-1, which is an alternative favored $\gamma$-ray production
site located at $\gtrsim$120~pc from the nucleus, remained quiescent in terms of
its flux density and structure. These results in the radio bands strongly suggest
that the VHE $\gamma$-ray activity in 2012 originates in the jet base within
0.03~pc or 56~Schwarzschild radii (the VERA spatial resolution of 0.4~mas at
43~GHz) from the central supermassive black hole. We further conducted VERA
astrometry for the M~87 core at six epochs during the flaring period, and detected
core shifts between 22 and 43~GHz, a mean value of which is similar to that
measured in the previous astrometric measurements. We also discovered a clear
frequency-dependent evolution of the radio core flare at 43, 22 and 5~GHz; the
radio flux density increased more rapidly at higher frequencies with a larger
amplitude, and the light curves clearly showed a time-lag between the peaks at 22
and 43~GHz, the value of which is constrained to be within $\sim35-124$~days. This
indicates that a new radio-emitting component was created near the black hole in
the period of the VHE event, and then propagated outward with progressively
decreasing synchrotron opacity. By combining the obtained core shift and time-lag,
we estimated an apparent speed of the newborn component propagating through the
opaque region between the cores at 22 and 43~GHz. We derived a sub-luminal speed
(less than $\sim$0.2$c$) for this component. This value is significantly slower
than the super-luminal ($\sim$1.1$c$) features that appeared from the core during
the prominent VHE flaring event in 2008, suggesting that the stronger VHE activity
can be associated with the production of the higher Lorentz factor jet in M~87.
\end{abstract}


\section{Introduction}
Active galactic nuclei (AGNs) generate powerful relativistic jets which are
thought to be a consequence of the material accretion onto the supermassive black
holes. The nearby radio galaxy M~87 accompanies one of the best studied AGN jets,
and its proximity \citep[$D=16.7$~Mpc;][]{blakeslee2009} and brightness have
enabled intensive studies of this jet over decades from radio, optical and to
X-ray~at tens of parsec scale resolutions~\citep[e.g.,][]{owen1989, biretta1999,
harris2006}. Moreover, the inferred very massive black hole~\citep[$M_{\rm BH}
\simeq (3-6) \times 10^9~M_{\odot}$;][]{macchetto1997, gebhardt2009, walsh2013}
yields a linear resolution down to $1~{\rm milliarcsecond~(mas)}=0.08~{\rm
pc}=140$ Schwarzschild radii~$(R_{\rm s})$~\footnote{In this paper we adopt
$M_{\rm BH} = 6\times 10^9M_{\odot}$ along with \citet{hada2011, hada2012,
hada2013}, although we note that the exact value of $M_{\rm BH}$ in M~87 is still
controversial and should be carefully considered. One can rescale the values in
$R_{\rm s}$ unit in this paper by multiplying a factor of 2 if $M_{\rm BH}=3\times
10^9M_{\odot}$ is used.}, making this source a unique case to probe the
relativistic-jet formation at an unprecedented compact scale with
Very-Long-Baseline-Interferometer (VLBI) observations~\citep{junor1999,
kovalev2007, hada2011, doeleman2012, hada2013}.

With the advent of the new-generation Cherenkov telescope arrays, M~87 is now well
known to show $\gamma$-ray emission up to the very-high-energy (VHE; $E>100$~GeV)
regime, where this source often exhibits active flaring episodes. The location and
the physical processes of such emission have been a matter of debate over the past
years. To date, M~87 underwent three remarkable flaring events in 2005, 2008 and
2010. In 2005, a VHE flare~\citep{aharonian2006} was accompanied by the
radio-to-X-ray outbursts from HST-1, a violent knot located at a de-projected
distance of $\gtrsim$120~pc downstream of the
nucleus~\citep{harris2006}\footnote{The exact distance of HST-1 depends on the
viewing angle of the M~87 jet $i$. While \citet{ly2007} estimates $i\sim
30^{\circ}-45^{\circ}$ based on their sub-parsec radio jet study, the observed
superluminal motions of HST-1 up to 6\,$c$ (at optical wavelength) suggest a
smaller viewing angle close to $i \sim$15$^{\circ}$~\citep{biretta1999}. The
recent optical polarization study of HST-1 also suggests the trend of the smaller
viewing angle mentioned above~\citep{perlman2011}. A detailed discussion for the
viewing angle of M~87 is given in \citet{acciari2009}, who suggested a likely
range to be $i \sim 15^{\circ}-25^{\circ}$ by surveying the literature.}, with
the emergence of superluminal ($\sim$$4c$) radio
features~\citep{cheung2007}. These results led to the argument that HST-1 is
associated with the VHE $\gamma$-ray production~\citep[e.g.,][]{stawarz2006,
cheung2007, harris2008, harris2009}.  In the case of the 2008
event~\citep{acciari2009}, on the other hand, the \textit{Chandra X-ray
Observatory} detected an enhanced X-ray flux from the nucleus, while HST-1
maintained a comparatively constant flux. In addition, the VLBA observations at
43~GHz during the VHE activity detected a strong flux increase from the radio core
at the jet base. These facts provide the strong claim that the VHE flare
originates in the core~\citep{acciari2009}.  The third event occurring in 2010 is
rather elusive. Coincident with the VHE event, \textit{Chandra} again detected an
enhanced flux from the X-ray core~\citep{harris2011, abramowski2012}, and VLBA
observations also suggested a possible increase of the radio core
flux~\citep{hada2012}. However, \citet{giroletti2012} found the emergence of a
superluminal component in the HST-1 complex near the epoch of this event, which is
reminiscent of the 2005 case.

At high-energy (HE; 100~MeV$<E<$100~GeV) $\gamma$ rays, the \textit{Fermi} Large
Area Telescope (LAT) detects a faint, point-like $\gamma$-ray emission within the
central $\sim$20~kpc of the galaxy in 10 months of all-sky
survey~\citep{abdo2009}. Contrary to the variable VHE $\gamma$-rays, the emission
in the MeV/GeV regime appears to be stable (at least) on a timescale of about
10$\sim$30 days, although the same amplitude of fluctuations as in TeV would have
been difficult to detect given the lower significance of detection in this band.
The observed MeV/GeV spectrum seems to connect smoothly with the VHE spectrum at a
low (non-flaring) state~\citep{abdo2009}.

Recently, the VERITAS Collaboration has reported new VHE $\gamma$-ray activity
from M~87 in early 2012~\citep{beilicke2012}. While there were no remarkable
flares like those in the previous episodes \citep[where the peak fluxes reached
$\Phi_{\rm > 0.35TeV}\sim$(1--3)$\times 10^{-11}$
photons~cm$^{-2}$~s$^{-1}$;][]{abramowski2012}, the VHE flux in 2012 clearly
exhibits an elevated state at a level of $\sim$$9\sigma$ ($\Phi_{\rm >
0.35TeV}\sim$(0.2--0.3)$\times 10^{-11}$ photons~cm$^{-2}$~s$^{-1}$) over the
consecutive two months from February to March 2012. The observed flux is a factor
of $\sim$2 brighter than that in the neighboring quiescent periods. To understand
the nature and the origin of this event in more detail, it is crucial to explore
the contemporaneous status of M~87 in other wavebands, particularly where
higher-resolution instruments are available.

Here we report a multi-wavelength radio and MeV/GeV study of the M~87 jet during
this period using the VLBI Exploration Radio Astrometry (VERA), the European VLBI
Network (EVN), the Submillimeter Array (SMA) and the \textit{Fermi}-LAT. We
especially focus on the VLBI data in the radio bands. With VERA, we obtained a
high-angular-resolution, phase-referencing data set at 22 and 43~GHz with a dense
time interval during the VHE activity in 2012. With the supportive EVN monitoring,
we obtained a complementary data set at 5~GHz, which enables a high-sensitivity
imaging of the M~87 jet. A collective set of these radio data allows us to probe
the detailed physical status and structural evolutions of M~87 by pinpointing the
candidate sites of the $\gamma$-ray emission i.e., the core and HST-1. In
Section~2, we present the radio data analyses for VERA, EVN and SMA. In Section~3,
we describe the selection and analysis of the \textit{Fermi}-LAT data. We then
describe our results and discussion in Section 4 and 5, respectively. In the final
section, we will summarize the paper. In the present paper, the spectral index
$\alpha$ is defined as $S_{\rm \nu} \propto \nu^{+\alpha}$.

\section{Radio data: observations and reduction}

\subsection{VERA dedicated sessions at 22 and 43~GHz (Mode-A)}
On 2012 February 24, March 16, 21, April 5, 27 and May 17, we made dedicated
observations of M~87 with VERA at 22.3 and 43.1~GHz (hereafter we term these
sessions as \textit{Mode-A}). All four of the VERA stations participated in good
weather conditions. Left-handed circular polarization was received and sampled
with a 2-bit quantization using the VERA digital filter unit. We employed
dual-beam observations to allow astrometric measurements; M~87 and a nearby radio
source M~84 (separated by 1.\hspace{-.3em}$^{\circ}$5 on the sky from M~87) were
observed simultaneously. The data were recorded at a rate of 1024~Mbps (a total
bandwidth of (128+128)~MHz), where each of the two 128~MHz-wide subbands was
allocated for each source. In an effort to reduce systematic errors and to achieve
similar $uv$ coverages between the two frequencies, each frequency was alternated
in turn every 80 minutes. At each epoch, the total on-source time results in about
3 hours at each frequency.

The initial data calibration was performed with the Astronomical Image Processing
System (AIPS) developed at the National Radio Astronomy Observatory (NRAO). First,
a-priori amplitude calibration was applied using the measured system noise
temperature and the elevation-gain curve of each antenna. We then calibrated the
amplitude part of bandpass characteristics at each station using the
auto-correlation data. Next, we employed the calibration of the fringe phases in
the following processes; first, we recalculated delay-tracking solutions for the
correlated data using an improved delay-tracking model. Throughout the data
analysis, we adopt the delay-tracking center as $\alpha_{\rm J2000}={\rm 12^{\rm
h}30^{\rm m}49.\hspace{-.3em}^{\rm s}4233830}$ and $\delta_{\rm J2000}={\rm
12^{\circ}23^{\prime}28.\hspace{-.3em}^{\prime\prime}043840}$ for M~87, while
$\alpha_{\rm J2000}={\rm 12^{h}25^{m}03.\hspace{-.3em}^{s}7433330}$ and
$\delta_{\rm J2000}={\rm
12^{\circ}53^{\prime}13.\hspace{-.3em}^{\prime\prime}139330}$ for M~84. The
delay-tracking solutions include delay contributions from the atmosphere, which
are estimated using the global positioning system data~\citep{honma2007}. Next, we
calibrated the differences due to instrumental delays between the two signal paths
in the dual beam, which are measured using artificial noise signals injected at
the same time into the two receivers~\citep{honma2008}. After that, we performed
fringe-fitting to remove residual delays, rates and phases. Because M~87 is much
brighter than M~84, we chose M~87 as a phase calibrator and transferred the
derived fringe solutions to the data of M~84. A fringe-fitting on M~87 was
conducted assuming a point source model. Deviations of the phase/gain solutions
from the point source model for M~87 were further corrected using the
self-calibrated images, the production process of which is described
below. Finally, the corrected phase and gain solutions were applied to M~84 and
its phase-referenced images were created, in which the relative position of M~84
with respect to M~87 is conserved. We describe the astrometry results in Section
4.3.

\subsection{VERA GENJI sessions at 22~GHz (Mode-B)}
In addition, VERA frequently observed M~87 at 22.2~GHz in the framework of the
\textit{GENJI} (\textit{Gamma-Ray Emitting Notable AGN Monitoring by Japanese
VLBI}) program~\citep{nagai2013}. This is a dense monitoring program of several
bright $\gamma$-ray AGN jets including M~87 with VERA starting from October 2010,
which takes advantage of the usual calibrator time for VERA's Galactic maser
astrometry observations. Between September 2011 and September 2012, M~87 was
routinely monitored with the \textit{GENJI} mode typically every 2--3 weeks
(hereafter we term these sessions as \textit{Mode-B}). Some of the data were not
useful because of the lack of one or two stations, or due to only one or two scans
available, preventing us from making reliable images and flux density
measurements. Then, we selected the data in which all four of the VERA stations
participated and at least three or four scans are available at different hour
angles. Through these selection criteria, we eventually obtained the
\textit{Mode-B} data at 18 epochs in total during this period. Total on-source
time for this mode is typically $\sim$30 minutes with an allocated bandwidth of
16~MHz. For more detailed descriptions about the data analysis of the
\textit{GENJI} data, see \citet{nagai2013}.

For all of the \textit{Mode-A} and \textit{Mode-B} data sets, images were created
in DIFMAP software with iterative CLEAN and phase/amplitude self-calibration. The
resultant amplitude corrections remained stable within $\sim$10\% over time at
each station, the level of which is consistent with a typical accuracy of the
system temperature measurements for VERA. We thus adopt 10\% error for the
amplitude uncertainty for the VERA data. The resultant synthesized beams with the
naturally-weighted scheme were typically $1.3\times 0.8$~mas in a position angle
(P.A.)$= -40^{\circ}$ and $0.65 \times 0.40$~mas in P.A.$=-40^{\circ}$ at 22 and
43~GHz, respectively. Comparing image qualities at 22~GHz, image rms levels
achieved by \textit{Mode-A} are twice (or more) as good as those obtained by
\textit{Mode-B} mainly because of the longer integration time and better
$uv$-coverage. We summarize the information about the VERA data in Table~1.

\begin{table}[htbp]
  \centering
 \begin{minipage}[t]{0.5\columnwidth}
  \caption{VERA and EVN observations of M~87}
  \medskip
    \small \begin{tabular}{cccccc}
    \hline
    \hline
    Array & Date  & Date  & Beam  & rms   & Mode \\
          &       & (MJD) & (mas$\times$mas, deg) & $\left(\frac{\rm mJy}{\rm beam}\right)$ &  \\
          &       &       &  (a)  &  (b)  &  (c)  \\
    \hline
    43GHz & 2012 Feb 24 & 55981 & $0.65\times0.41$, $-41$ & 1.9 & A \\
    VERA  & 2012 Mar 16 & 56002 & $0.63\times0.40$, $-40$ & 2.8 & A \\
          & 2012 Mar 21 & 56007 & $0.68\times0.44$, $-39$ & 1.2 & A \\
          & 2012 Apr 05 & 56022 & $0.65\times0.43$, $-43$ & 1.9 & A \\
          & 2012 Apr 27 & 56044 & $0.68\times0.44$, $-46$ & 1.8 & A \\
          & 2012 May 17 & 56064 & $0.65\times0.45$, $-45$ & 1.3 & A \\
    \hline
    22GHz & 2011 Sep 03 & 55808 & $1.32\times0.80$, $-32$ & 5.5 & B \\
    VERA  & 2011 Sep 07 & 55811 & $1.41\times0.77$, $-35$ & 2.9 & B \\
          & 2011 Sep 09 & 55813 & $1.34\times0.77$, $-35$ & 4.1 & B \\
          & 2011 Oct 15 & 55849 & $1.41\times0.80$, $-35$ & 2.7 & B \\
          & 2011 Nov 05 & 55870 & $1.31\times0.76$, $-37$ & 4.2 & B \\
          & 2011 Dec 04 & 55899 & $1.39\times0.84$, $-38$ & 2.3 & B \\
          & 2011 Dec 20 & 55915 & $1.41\times0.90$, $-18$ & 2.9 & B \\
          & 2012 Jan 23 & 55949 & $1.29\times0.87$, $-43$ & 1.6 & B \\
          & 2012 Feb 24 & 55981 & $1.40\times0.76$, $-40$ & 1.1 & A \\
          & 2012 Feb 27 & 55984 & $1.27\times0.89$, $-50$ & 1.8 & B \\
          & 2012 Feb 28 & 55985 & $1.57\times0.70$, $-43$ & 5.2 & B \\
          & 2012 Mar 16 & 56002 & $1.33\times0.84$, $-45$ & 2.7 & A \\
          & 2012 Mar 21 & 56007 & $1.40\times0.82$, $-36$ & 1.1 & A \\
          & 2012 Apr 01 & 56018 & $1.25\times0.96$, $-35$ & 2.0 & B \\
          & 2012 Apr 05 & 56022 & $1.39\times0.82$, $-40$ & 1.6 & A \\
          & 2012 Apr 19 & 56036 & $1.24\times0.85$, $-44$ & 2.7 & B \\
          & 2012 Apr 23 & 56040 & $1.50\times0.82$, $-25$ & 4.4 & B \\
          & 2012 Apr 27 & 56044 & $1.44\times0.84$, $-40$ & 1.5 & A \\
          & 2012 May 17 & 56064 & $1.36\times0.84$, $-40$ & 1.7 & A \\
          & 2012 May 23 & 56070 & $1.43\times0.78$, $-44$ & 2.6 & B \\
          & 2012 Jun 21 & 56099 & $1.29\times0.83$, $-53$ & 8.1 & B \\
          & 2012 Aug 29 & 56168 & $1.32\times0.96$, $-58$ & 8.2 & B \\
          & 2012 Sep 13 & 56183 & $1.34\times1.04$, $-55$ & 5.2 & B \\
          & 2012 Sep 23 & 56193 & $1.41\times0.78$, $-32$ & 3.9 & B \\
    \hline
    5GHz  & 2011 Mar 09 & 55629 & $3.05\times2.10$, $-67$ & 0.044 & -- \\
    EVN   & 2011 Apr 12 & 55663 & $2.71\times1.53$, $86$  & 0.092 & -- \\
          & 2011 Jun 02 & 55714 & $1.42\times1.05$, $-11$ & 0.135 & -- \\
          & 2011 Aug 25 & 55798 & $3.09\times1.29$, $-73$ & 0.133 & -- \\
          & 2011 Oct 17 & 55851 & $6.21\times2.29$, $-48$ & 0.099 & -- \\
          & 2012 Jan 10 & 55936 & $1.63\times1.16$, $-72$ & 0.070 & -- \\
          & 2012 Mar 20 & 56006 & $1.83\times1.67$, $66$  & 0.054 & -- \\
          & 2012 Jun 19 & 56097 & $2.82\times1.37$, $-89$ & 0.094 & -- \\
          & 2012 Oct 09 & 56209 & $3.27\times2.29$, $-85$ & 0.078 & -- \\
    \hline
    \end{tabular}
 \end{minipage}
 \label{tab:addlabel} \newline \textbf{Notes.} Columns: (a) size of a synthesized
 beam with a naturally-weighted scheme; (b) rms image noise level at a region far
 from the core with the beam (a); (c) VERA observation mode for each epoch. ``A''
 represents the dedicated mode, while ``B'' indicates the GENJI mode.
\end{table}

\subsection{EVN data at 5~GHz}
We observed M~87 with EVN at 9 epochs between March 2011 and October 2012 as a
continuation of our M~87/HST-1 monitoring project starting from mid
2009~\citep{giovannini2011, giroletti2012}. The observation at each epoch
typically lasted 4--8 hours, and the longest baselines were achieved from European
stations to Shanghai, Arecibo and/or Hartebeesthoek. Overall, the data quality is
adequate to warrant good signal-to-noise detections of the source structure.  For
all observations, the sky frequency was centered at 5.013~GHz and divided into
eight sub-bands separated by 16~MHz each for an aggregate bit rate of 1024
Mbps. The data were correlated in real time at the Joint Institute for VLBI in
Europe (JIVE) and automated data flagging and initial amplitude/phase calibration
were also carried out at JIVE using dedicated pipeline scripts. The data were
finally averaged in frequency within each IF, but individual IFs were kept
separate to minimize bandwidth smearing. Similarly, the data were time-averaged to
only 8 seconds to minimize time smearing. We produced final images in DIFMAP after
several cycles of phase and amplitude self-calibration. Various weighting schemes
were applied to the data to improve resolution in the core region and enhance the
fainter emission in the HST-1 region. For the core region, uniform weights or
natural weights were used without $uv$-tapering, which achieve angular resolutions
typically between 1--3~mas. For the HST-1 region, we used natural weights with a
Gaussian $uv$-tapering (typically a factor of 0.3 at a radius of 50~M$\lambda$,
which results in a beam size between 5--10~mas. Image rms noise levels estimated
in a region far from the core are typically 0.1~mJy/beam or less with the natural
weights. Regarding uncertainties on flux density measurements, we adopted 10\%
error for core flux density based on a typical EVN amplitude calibration
accuracy. For HST-1, we conservatively assumed 30\% uncertainty because analyses
of the HST-1 region could contain a larger amount of deconvolution error due to
its weak and extended nature. Details of the EVN data are also summarized in
Table~1.

\subsection{SMA data at 230~GHz}

The 230~GHz (1.3 mm) light curve was obtained at the SMA near the summit of Mauna
Kea (Hawaii). M~87 is included in an ongoing monitoring program at the SMA to
determine the flux densities of compact extragalactic radio sources that can be
used as calibrators at mm wavelengths~\citep{gurwell2007}.  Observations of
available potential calibrators are observed for 3 to 5 minutes, and the measured
source signal strength calibrated against known standards, typically solar system
objects (Titan, Uranus, Neptune, or Callisto). Data from this program are updated
regularly and are available at the SMA
website\footnote{http://sma1.sma.hawaii.edu/callist/callist.html}. Most of the SMA
measurements were obtained in its compact array configuration and the typical
angular resolution is about $\sim$3 arcseconds.

\section{Fermi-LAT Data: Selection and Analysis}
\label{FermiData}

The \textit{Fermi}-LAT is a pair-conversion telescope operating from 20 MeV to $>$
300 GeV. Further details about the \textit{Fermi}-LAT are given in
\citet{atwood09}. The LAT data reported in this paper were collected from 2011
February 1 (MJD 55593) to 2012 September 30 (MJD 56200). During this time, the
\textit{Fermi} observatory operated almost entirely in survey mode. The analysis
was performed with the \texttt{ScienceTools} software package version
v9r32p5\footnote{http://fermi.gsfc.nasa.gov/ssc/data/analysis/documentation/}.
The LAT data were extracted within a $10^{\circ}$ region of interest centred at
the radio location of M~87. Only events belonging to the `Source' class were
used. The time intervals when the rocking angle of the LAT was greater than
52$^{\circ}$ were rejected. In addition, a cut on the zenith angle ($<
100^{\circ}$) was applied to reduce contamination from the Earth limb $\gamma$
rays, which are produced by cosmic rays interacting with the upper atmosphere.
The spectral analysis was performed with the instrument response functions
\texttt{P7REP\_SOURCE\_V15}\footnote{http://www.slac.stanford.edu/exp/glast/canda/lat\_Peformance.htm}
using an unbinned maximum-likelihood method implemented in the Science tool
\texttt{gtlike}. A Galactic diffuse emission model and isotropic component, which
is the sum of an extragalactic and residual cosmic ray background, were used to
model the
background\footnote{http://fermi.gsfc.nasa.gov/ssc/data/access/lat/BackgroundModels.html}. The
normalizations of both components in the background model were allowed to vary
freely during the spectral fitting.

We evaluated the significance of the $\gamma$-ray signal from the sources by means
of the maximum-likelihood test statistic TS = 2$\Delta$log(likelihood) between
models with and without a point source at the position of M~87
\citep{mattox96}. The source model used in \texttt{gtlike} includes all of the
point sources from the second \textit{Fermi}-LAT catalog \citep[2FGL;][]{nolan12}
that fall within $15^{\circ}$ of the source. The spectra of these sources were
parametrized by power-law functions, except for 2FGL\,J1224.9+2122 (4C\,21.35) and
2FGL\,J1229.1+0202 (3C\,273), for which we used a log-parabola as in the 2FGL
catalogue. A first maximum-likelihood analysis was performed to remove from the
model the sources having TS $<$ 25 and/or the predicted number of counts based on
the fitted model $N_{pred} < 3 $. A second maximum-likelihood analysis was
performed on the updated source model. In the fitting procedure, the normalization
factors and the photon indices of the sources lying within 10$^{\circ}$ of M~87
were left as free parameters. For the sources located between 10$^{\circ}$ and
15$^{\circ}$, we kept the normalization and the photon index fixed to the values
from the 2FGL catalogue.

Integrating over the period from 2011 February 1 to 2012 September 30 (MJD
55593--56200), the fit with a power-law model in the 0.1--100 GeV energy range
results in a TS = 134, with an integrated average flux of (2.22 $\pm$ 0.43)
$\times$10$^{-8}$ ph cm$^{-2}$ s$^{-1}$ and a photon index of $\Gamma$ = 2.25
$\pm$ 0.10. Taking into account the detection significance over the whole analysed
period, we produced the $\gamma$-ray light curves with 1-month and 2-month time
bins. This choice of binning is compatible with those adopted in the previous M~87
studies with LAT data~\citep{abdo2009, abramowski2012}, and also reasonable for a
comparison with the observed month-scale VHE activity in 2012. For each time bin,
the spectral parameters for M~87 and for all the sources within 10$^{\circ}$ from
it were frozen to the value resulting from the likelihood analysis over the entire
period. In the light curve with the 2-month time bins, if TS $<$ 10, 2$\sigma$
upper limits were evaluated, while only bins with TS $>$ 10 are selected in the
light curve with the 1-month time bins. We describe the results of the LAT light
curves in Section~4.2.

Dividing the 1-month bins with higher flux in 5-day sub-bins, the highest flux of
(10.4$\pm$4.8)$\times$10$^{-8}$ and (8.6$\pm$3.4)$\times$10$^{-8}$ ph cm$^{-2}$
s$^{-1}$ was detected on 2011 October 12-16 and 2012 January 16-20, respectively
(these sub-bin data also show TS$>$10). By means of the \texttt{gtsrcprob} tool,
we estimated that the highest energy photon emitted from M~87 (with probability
$>$ 90\% of being associated with the source) was observed by LAT on 2011 April 7,
at a distance of 0.09$^{\circ}$ from the source and with an energy of 254.0 GeV,
extending into the VHE range.

\begin{figure}[htbp]
 \centering
 \includegraphics[angle=0,width=1.0\textwidth]{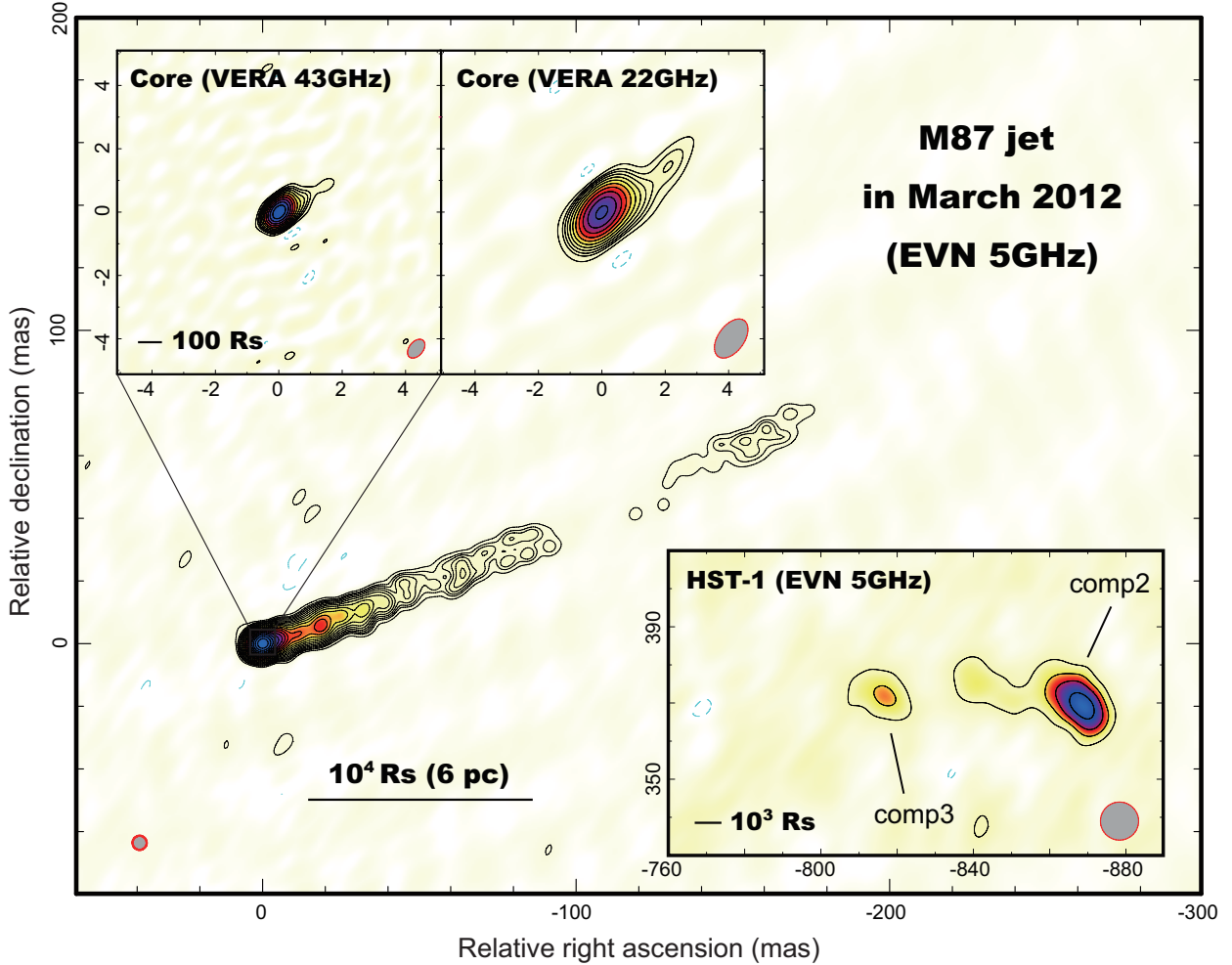}
 \caption{Montage of VLBI images of the M 87 jet during the elevated VHE
 $\gamma$-ray state in 2012 March. The main (global) image was obtained with EVN
 at 5~GHz (on March 20). The bottom right inset indicates a close-up view toward
 the HST-1 region. The nomenclatures of the two main features (comp2 and comp3)
 are based on \citet{giroletti2012}. The upper left two insets indicate VERA
 images for the core at 22 and 43~GHz (on March 21). For the EVN images, a
 circular Gaussian beam of 4.5~mas is used for the global structure (bottom left
 in the image), while a 10.0~mas circular beam is applied for HST-1 (bottom right
 in the HST-1 image). For the VERA images, the beam sizes are and
 $1.40\times0.82$~mas in P.A. $-36^{\circ}$ and $0.68\times0.44$~mas in
 P.A. $-39^{\circ}$ at 22 and 43~GHz, respectively (bottom right in each VERA
 image). For each image, contours start from $-$1, 1, 2, 2$^{3/2}$, 4,
 2$^{5/2}$... times 3$\sigma$ image rms levels (where $1\sigma$ are set to be
 1.1~mJy~beam$^{-1}$ (22-GHz image), 1.2~mJy~beam$^{-1}$ (43-GHz image),
 0.48~mJy~beam$^{-1}$ (EVN inner jet image) and 0.08~mJy~beam$^{-1}$ (EVN HST-1
 image)) and increase by factors of 2$^{1/2}$.}  \label{fig:}
\end{figure}

\section{Results}
\subsection{Milliarcsecond scale images}
Figure~1 shows VLBI images for M~87 during the elevated VHE state in March
2012. The large scale jet structure down to HST-1 was clearly detected with EVN,
while VERA resolved the innermost region.

The obtained EVN jet images within several hundreds of mas from the nucleus are in
good agreement with the well-known characteristics for this source; the jet is
described by the bright radio core at the base with the edge-brightened collimated
region~\citep{junor1999, ly2004, dodson2006, ly2007, asada2012, hada2013}. On the
other hand, the VERA array is less sensitive to the extended emission (i.e.,
mostly resolved out) due to the lack of short baselines ($>60$~M$\lambda$ and
$>120$~M$\lambda$ at 22 and 43~GHz respectively), so the detectable emission is
generally concentrated within the central region (i.e., a few mas). Nevertheless,
some of brighter features downstream of the core were indeed consistently detected
at many of the analyzed epochs. Our preliminary kinematic measurements of these
features with VERA suggest an apparent motion around $\sim$0.4\,$c$ relative to
the radio core~\citep{hada2013b}, which is similar to the value obtained in a
previous VLBA study~\citep{ly2007}. However, these features have already existed
from mid 2011 and we did not find any notable correlations with the 2012 VHE
activity. Thus, here we do not describe further details on these features and the
more dedicated treatment will be reported in a forthcoming paper.

In terms of the HST-1 region, we detected the detailed substructure in all of the
analyzed EVN data, the characteristics of which are overall consistent with our
previous study~\citep{giroletti2012}. We confirmed that the main emission features
of HST-1 are still moving constantly at a superluminal speed of $\sim$4$c$. While
in \citet{giroletti2012} we identified three main features (termed as comp1, comp2
and comp3 from the downstream side), in Figure 1 we can only see two main
components. These features correspond to comp2 (the downstream, brighter one) and
comp3 (the upstream, weaker one), respectively. The comp1, which was identified as
the outermost feature in \citet{giroletti2012}, faded gradually and became below
the noise limit at the beginning of 2012. While the emergence of new components
from the upstream edge of HST-1 (around $({\rm RA}, {\rm Dec})=(-780, 350)$~mas
from the core) were seen during the VHE events in 2005 and 2010, we did not find
such a structural variation during the 2012 event.

\begin{figure}[htbp]
 \centering \includegraphics[angle=0,width=0.75\columnwidth]{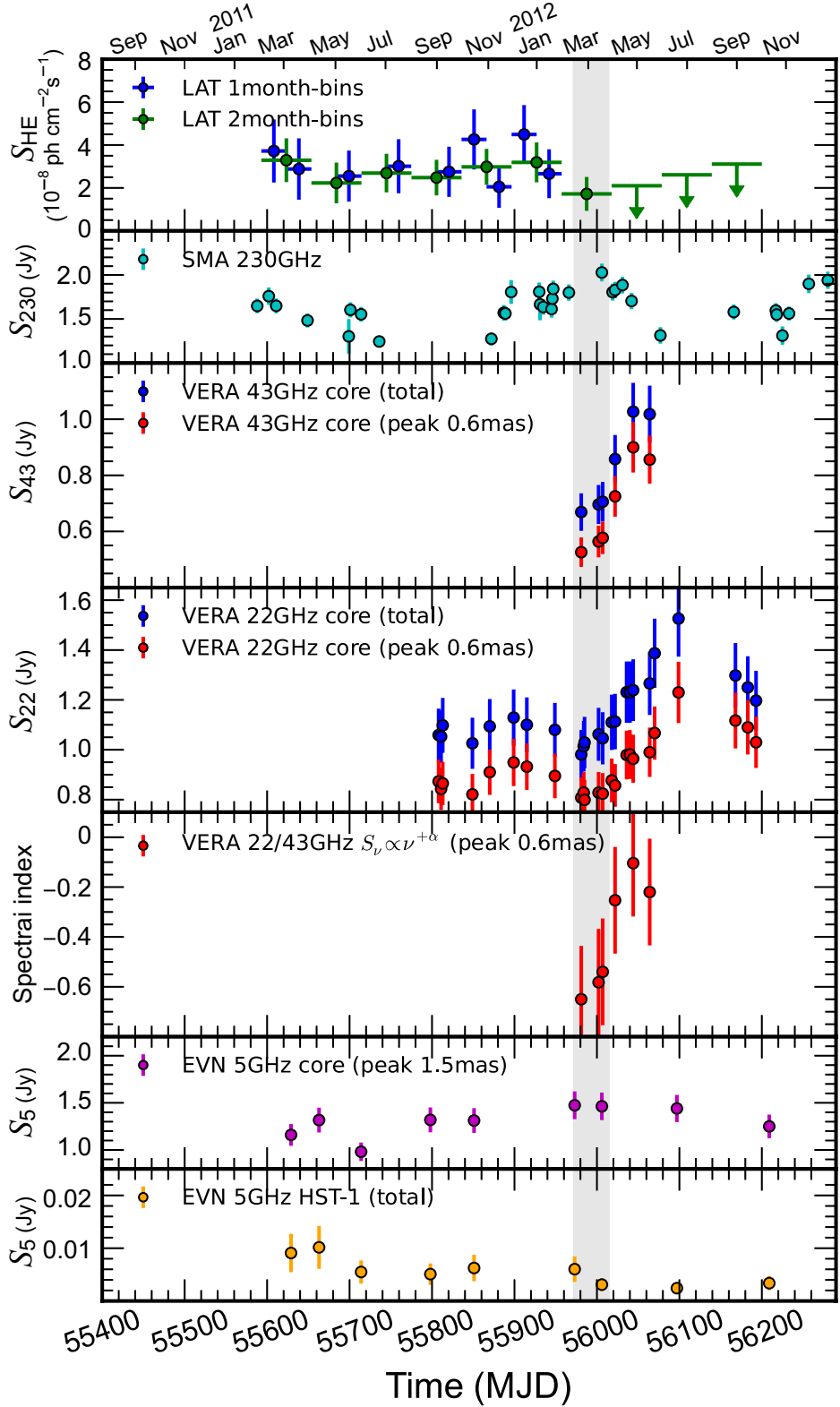}
 \caption{Multi-wavelength light curves of M~87 between February 2011 and December
 2012. The vertical shaded area over the plots indicates a period of elevated VHE
 emission reported by \citet{beilicke2012}. See the text for more detailed
 description for the each panel of the light curve.} \label{fig:}
\end{figure}

\begin{figure}[htbp]
 \centering \includegraphics[angle=0,width=0.7\columnwidth]{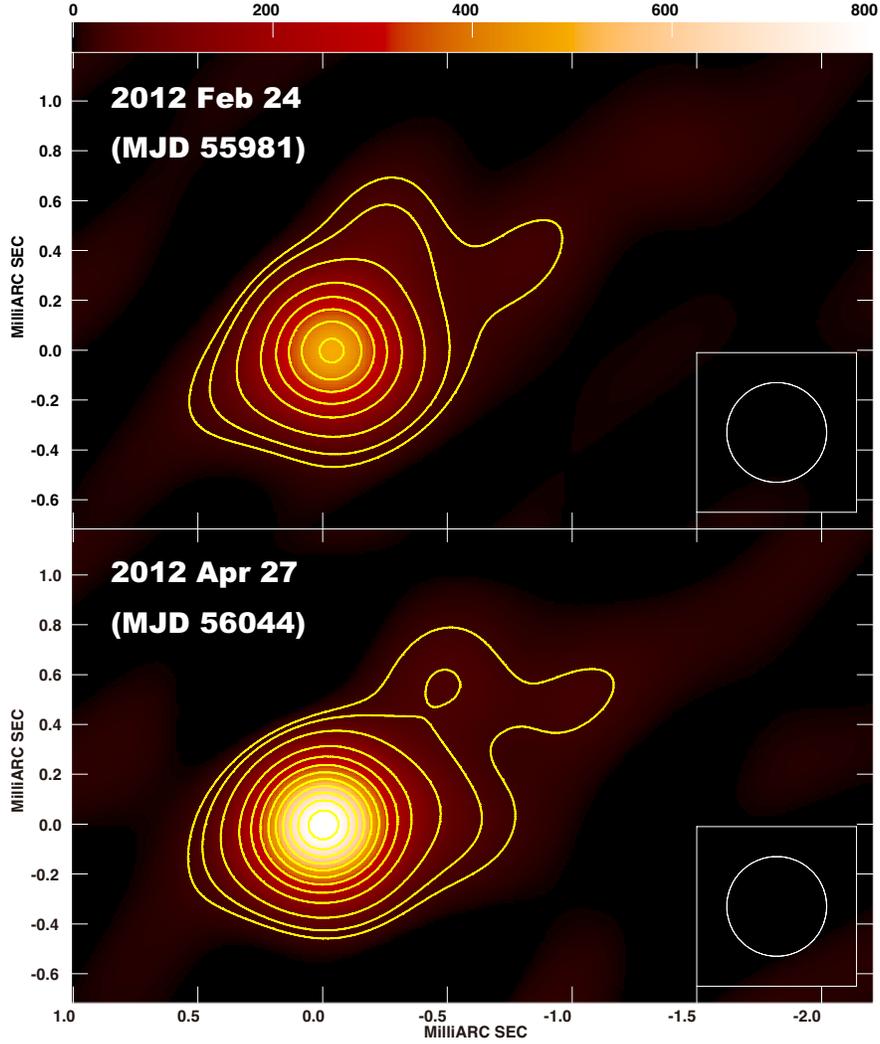}
 \caption{VERA 43-GHz images of the M~87 jet during the elevated VHE state in
 2012. The images are convolved with a 0.4 mas diameter circular Gaussian beam
 (shown at the bottom-right corner of each image), the size of which is close to
 the FWHM of the minor axis of the synthesized beam. The range of intensities is
 represented both in color and with contours at levels of
 800~mJy~beam$^{-1}$$\times$ (0.01, 0.05, 0.1, 0.2, 0.3, 0.4, 0.5, 0.6, 0.7, 0.8,
 0.9, 1.0).} \label{fig:}
\end{figure}

\begin{table}[htbp]
 \begin{minipage}[t]{1.0\textwidth}
   \centering
  \caption{Flux density measurements with VERA at 22 and 43~GHz}
    \begin{tabular*}{1.0\textwidth}{@{\extracolsep{\fill}}ccccccc}
    \hline
    \hline
    Date  & Date  & $S_{\rm 22, tot}$ & $S_{\rm 22, peak}$ & $S_{\rm 43, tot}$
     &$S_{\rm 43, peak}$ & $\alpha_{\rm 22-43, peak}$ \\
          & (MJD) & (mJy) & $\left(\frac{\rm mJy}{\rm beam}\right)$ & (mJy) &
			 $\left(\frac{\rm mJy}{\rm beam}\right)$ &  \\
          &       & (a)  & (b) & (c)  &  (d)  &  (e) \\
    \hline
    2011 Sep 03 & 55808 & $1059\pm106$  & $873\pm87$   &       &   &   \\
    2011 Sep 07 & 55811 & $1053\pm105$  & $843\pm84$   &       &   &   \\
    2011 Sep 09 & 55813 & $1098\pm110$  & $865\pm87$   &       &   &   \\
    2011 Oct 15 & 55849 & $1026\pm103$  & $821\pm82$   &       &   &   \\
    2011 Nov 05 & 55870 & $1094\pm109$  & $910\pm91$   &       &   &   \\
    2011 Dec 04 & 55899 & $1129\pm113$  & $949\pm95$   &       &   &   \\
    2011 Dec 20 & 55915 & $1100\pm110$  & $932\pm93$   &       &   &   \\
    2012 Jan 23 & 55949 & $1080\pm108$  & $895\pm90$   &       &   &   \\
    2012 Feb 24 & 55981 & $981\pm98$    & $808\pm81$   & $669\pm67$   & $526\pm53$
			 & $-0.65\pm0.21$   \\
    2012 Feb 27 & 55984 & $1014\pm101$  & $828\pm83$   &       &   &   \\
    2012 Feb 28 & 55985 & $1029\pm110$  & $799\pm80$   &       &   &   \\
    2012 Mar 16 & 56002 & $1062\pm106$  & $828\pm83$   & $696\pm70$   & $564\pm56$
			 & $-0.58\pm0.21$   \\
    2012 Mar 21 & 56007 & $1046\pm105$  & $824\pm82$   & $706\pm71$   & $577\pm58$
			 & $-0.54\pm0.21$   \\
    2012 Apr 01 & 56018 & $1110\pm111$  & $877\pm88$   &       &   &   \\
    2012 Apr 05 & 56022 & $1113\pm111$  & $857\pm86$   & $858\pm86$   & $725\pm73$
			 & $-0.25\pm0.21$   \\
    2012 Apr 19 & 56036 & $1231\pm123$  & $979\pm98$   &       &   &   \\
    2012 Apr 23 & 56040 & $1204\pm120$  & $981\pm98$  &       &   &   \\
    2012 Apr 27 & 56044 & $1239\pm124$  & $964\pm96$   & $1027\pm103$  &
			 $900\pm90$  & $-0.10\pm0.21$   \\
    2012 May 17 & 56064 & $1266\pm127$  & $990\pm99$  & $1018\pm102$  & $856\pm86$
			 & $-0.22\pm0.21$   \\
    2012 May 23 & 56070 & $1387\pm139$  & $1067\pm107$  &       &   &   \\
    2012 Jun 21 & 56099 & $1526\pm153$  & $1230\pm123$  &       &   &   \\
    2012 Aug 29 & 56168 & $1298\pm130$  & $1117\pm112$  &       &   &   \\
    2012 Sep 13 & 56183 & $1263\pm126$  & $1090\pm109$  &       &   &   \\
    2012 Sep 23 & 56193 & $1197\pm120$  & $1072\pm107$  &       &   &   \\
    \hline
    \end{tabular*}
 \end{minipage}
  \label{tab:addlabel} Notes: (a) and (c) Integrated flux density of VERA images
 at 22~GHz and 43~GHz, respectively: (b) and (d) Peak flux density of VERA at
 22~GHz and 43~GHz (with a 0.6-mas-diameter circular Gaussian convolving beam),
 respectively: (e) Spectral index calculated by $\log({S_{\rm 22, peak}/S_{\rm 43,
 peak}})/\log(22.291/43.125)$.
\end{table}

\begin{table}[htbp]
 \begin{minipage}[t]{1.0\columnwidth}
  \centering \caption{Flux density measurements with EVN at 5~GHz}
    \begin{tabular*}{1.0\columnwidth}{@{\extracolsep{\fill}}ccccccc}
    \hline
    \hline
    Date  & Date  & $S_{\rm 5,core}$ & $S_{\rm 5,HST-1}$ \\
          & (MJD) & $\left(\frac{\rm mJy}{\rm beam}\right)$ & (mJy)    \\
          &       &   (a)  & (b)           \\
    \hline
    2011 Mar 09 & 55629 & $1160\pm116$  & $9.1\pm2.7$   \\
    2011 Apr 12 & 55663 & $1317\pm132$  & $10.2\pm3.6$  \\
    2011 Jun 02 & 55714 & $981\pm98$   & $5.5\pm1.7$   \\
    2011 Aug 25 & 55798 & $1319\pm132$  & $5.1\pm1.5$   \\
    2011 Oct 17 & 55851 & $1312\pm131$  & $6.3\pm1.9$   \\
    2012 Jan 10 & 55936 & $1474\pm147$  & $6.0\pm1.8$   \\
    2012 Mar 20 & 56006 & $1463\pm146$  & $3.1\pm0.9$   \\
    2012 Jun 19 & 56097 & $1440\pm144$  & $2.4\pm0.7$   \\
    2012 Oct 09 & 56209 & $1250\pm125$  & $3.4\pm1.0$   \\
    \hline
    \end{tabular*}
  \end{minipage}
  \label{tab:addlabel} Notes: (a) peak flux density of the core when convolved by
 a 1.5-mas-diameter circular Gaussian beam: (b) integrated flux density of the
 HST-1 region.
\end{table}

\subsection{MWL Light Curves}
In Figure~2, we show a combined set of light curves of M~87 from radio to MeV/GeV
$\gamma$-ray between MJD~55400 and MJD~56280. The vertical shaded area over the
plots indicates a period between February 2012 and March 2012, where an elevated
VHE state is reported by VERITAS~\citep{beilicke2012}. For the LAT data, we show
the light curves produced with 1-month and 2-month time bins (including 2$\sigma$
upper limits for the 2-month time bins). For the SMA data, we plot a light curve
of the total flux densities. For the EVN 5-GHz data, we show the light curves for
both the HST-1 region (integrated flux densities) and the core (peak flux
densities when convolved by a 1.5-mas circular beam). In terms of the VERA data at
22/43~GHz, the light curves obtained by the \textit{Mode-A/B} are combined into a
single sequence. At each frequency, we provide both the integrated flux density
and the peak flux density when convolved with a 0.6-mas circular Gaussian
beam. Additionally, we plot an evolution of the spectral index for the peak flux
density between 22 and 43~GHz.

Thanks to the dense, complementary coverages of VERA and EVN, we revealed the
detailed evolutions of the radio light curves for both the core and HST-1. The
most remarkable finding in these plots is a strong enhancement of the radio core
flux density at VERA 22 and 43~GHz that starts from MJD~55981 (2012 February
24). While the timing of the radio peak is delayed, the onset of the radio flare
occurs coincidentally with the enhanced VHE activity. At 22~GHz, we further
detected a subsequent decay stage of the brightness at the last three epochs. Also
at 43~GHz, we detected possible saturation of the flux increase near the last
epoch. Meanwhile, the EVN monitoring confirmed a constant decrease of the HST-1
luminosity. We also checked the light curve for the peak flux density of HST-1
(with a 10-mas-diameter convolving beam), but the overall trend is the same as
that of the integrated flux density. We summarize the values of the measured flux
with VERA and EVN in Table~2 and Table~3.

To better describe the structure of the core region, we created VERA 43-GHz images
convolved with a circular Gaussian beam of 0.4-mas diameter (Figure~3), which are
$\sim$30\% super-resolution along the major axis of the synthesized beam.  These
images clearly indicate that the flux enhancement occurs within the central
spatial resolution of 0.4~mas, corresponding to a linear scale of 0.03 pc or
56~$R_{\rm s}$. During the period of the elevated VHE state, the SMA data at
230~GHz also appear to show a local maximum in its light curve, although we note
that its angular resolution is significantly large (generally $\sim$1--3
arcseconds, containing both the core and HST-1).

\begin{figure}[htbp]
 \centering
 \includegraphics[angle=0,width=0.75\columnwidth]{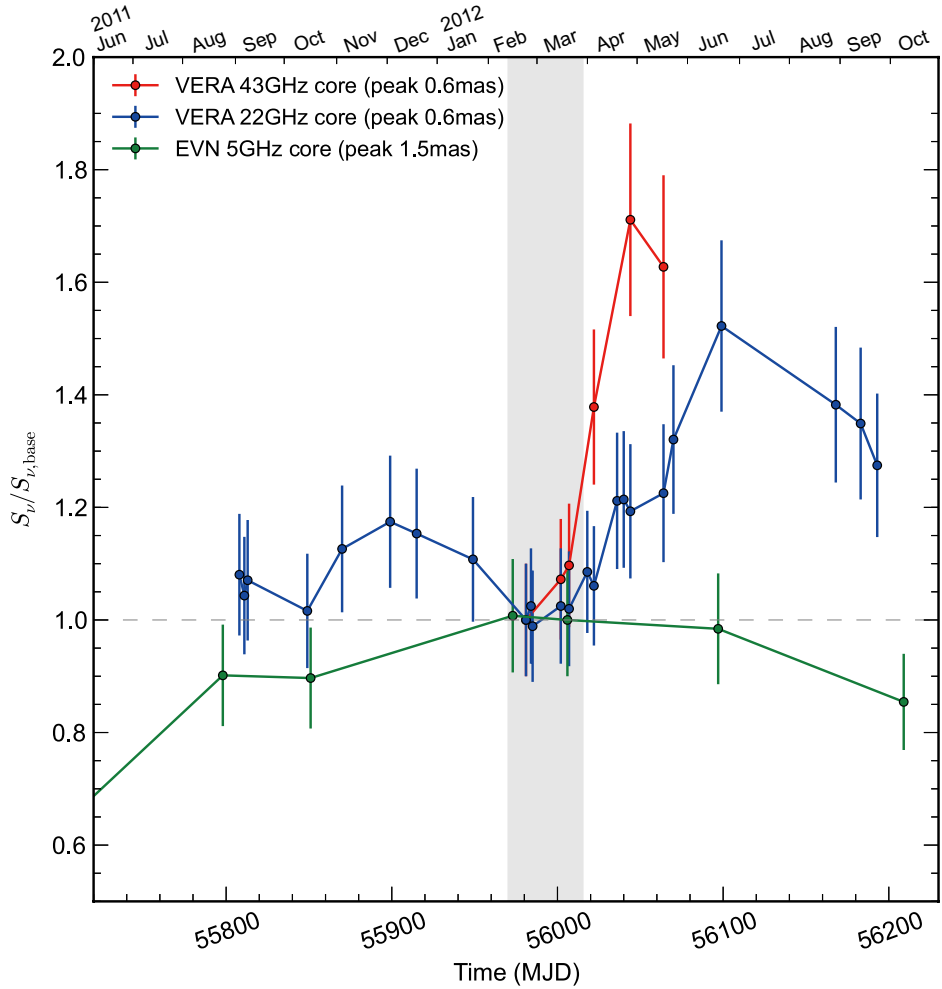}
 \caption{Normalized radio light curves of the M~87 core at 43 (peak 0.6 mas), 22
 (peak 0.6~mas) and 5~GHz (peak 1.5~mas). The light curves at 43, 22 and 5~GHz are
 normalized to the flux values on MJD 55981, MJD 55981 and MJD 55936,
 respectively. The vertical shaded area over the plots indicates a period of the
 elevated VHE state reported by \citet{beilicke2012}} \label{fig:}
\end{figure}

Another notable finding by our monitoring is a frequency-dependent evolution of
the radio core flare. Figure~4 summarizes a set of the radio light curves that are
normalized by the ``baseline level'' of the radio flare at each frequency (i.e.,
the flux values on MJD 55981 at 22/43~GHz and on MJD 55936 at 5~GHz, after which
the radio core starts to flare). This figure clearly indicates that the radio core
brightens more rapidly with a larger amplitude as frequency increases. At 43~GHz,
the flux density increased up to $\sim$70\% for the subsequent two months (between
MJD~55981 and MJD~56044) at an averaged rate of $\sim$35\%/month, and afterward
the growth seems to be saturated. On the other hand, the core flux density at
22-GHz progressively increased up to $\sim$50\% (at least) for the subsequent four
months (from MJD~55981 to MJD~56099) at a slower rate of
$\sim$12\%/month. Eventually, the spectral index of the radio core between the two
frequencies changed continuously during the flaring event (from $\alpha\sim-0.6$
to $\sim-0.1$ for the peak flux with a 0.6-mas beam; see Figure~2). At 5~GHz, by
contrast, the core remained virtually stable within the adopted error of
10\%. This is the first time that such a frequency-dependent nature of the radio
flare is clearly detected in the M~87 jet.

With regard to the MeV/GeV regime, the LAT light curves were stable up to February
2012 (both 1- and 2-month time bins), and we did not find any significant flux
enhancement during the period of the VHE activity. The observed HE flux level was
consistent with that seen in the earlier epochs~\citep{abdo2009, abramowski2012}
within the adopted uncertainty. After March 2012, however, no significant emission
was detected for the subsequent 6 months in the 1- and 2-month binned data,
suggesting a change in the HE state after the VHE event. To quantify this, we
analyzed the LAT data in two 6-month intervals, 2011 October-2012 March and 2012
April-September, following the same procedure described in Section~3. The fits
resulted in 0.1-100 GeV fluxes of $(2.6 \pm 0.5) \times 10^{-8}$ ph cm$^{-2}$
s$^{-1}$ (TS = 56) and $(1.1 \pm 0.5) \times 10^{-8}$ ph cm$^{-2}$ s$^{-1}$ (TS =
10), respectively. This indicates a decrease in the HE flux by a factor of $\sim
2$ after the VHE event, in agreement with the level of decrease observed at VHE in
2012 April-May~\citep{beilicke2012}.

\subsection{Astrometry of the core}

Since the VERA \textit{Mode-A} performed astrometry observations at both 22 and
43~GHz, these data enable us to examine the core shift effect during the VHE
activity in 2012. The core-shift effect is expected as a natural consequence of
the nuclear opacity effect if the radio core at each frequency corresponds to a
surface of synchrotron-self-absorption (SSA) $\tau_{\rm
ssa}(\nu)\sim1$~\citep{konigl1981, lobanov1998, hada2011}.

In Figure~5, we show the result of our VERA astrometry for the M~87 radio core at
22 and 43~GHz, which are measured relative to the radio core of M~84 at each
frequency. For each data, we defined the core position as follows. For M~87, we
convolved the self-calibrated images by a 0.6-mas/0.4-mas-circular Gaussian beam
at 22/43~GHz, the length of which is virtually equal to that of the minor axis of
synthesized beam at each frequency. Then, we determined a brightness centroid of
the image by fitting a Gaussian model with the AIPS task JMFIT. For M~84, we used
its phase-referenced images instead of self-calibrated ones. Regarding position
errors adopted for each data point in Figure~5, here we considered a
root-sum-square of individual errors from the peak-to-noise ratio of the M~84
phase-referenced image, tropospheric residuals, ionospheric residuals and
geometrical errors.

\begin{figure}[htbp]
 \centering \includegraphics[angle=0,width=0.75\columnwidth]{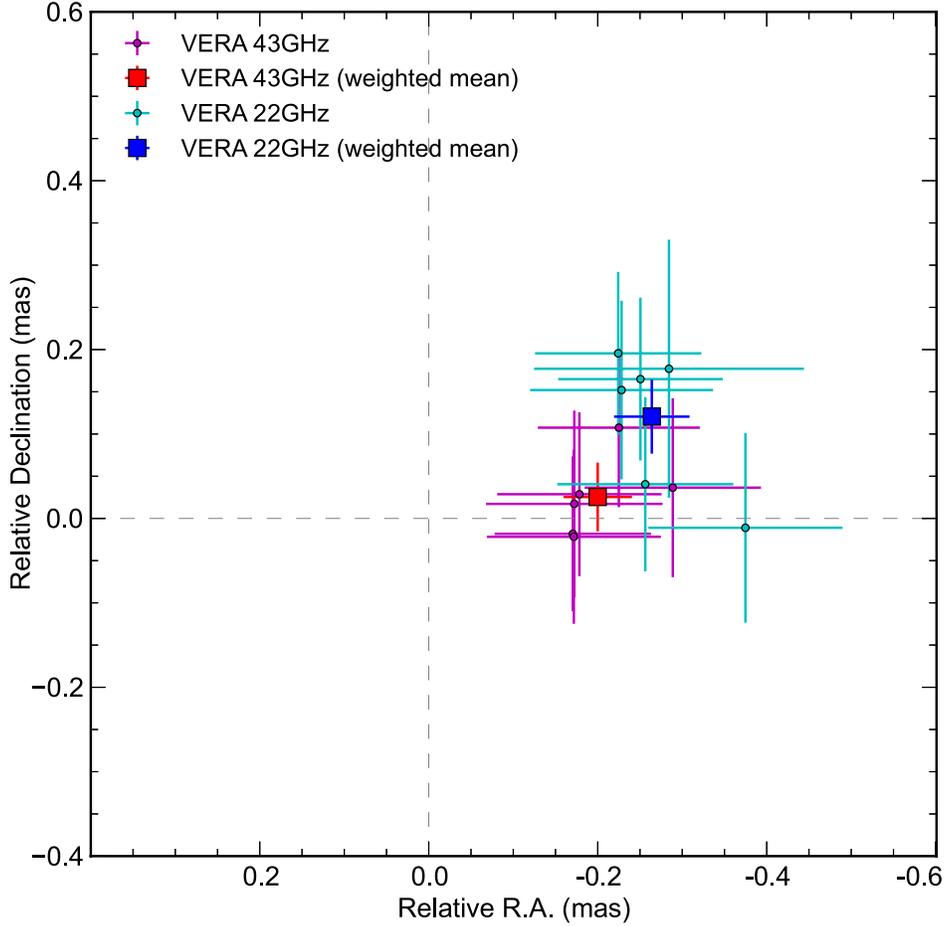}
 \caption{VERA astrometry of the M~87 core position at 22 and 43~GHz with respect
 to the core position of M~84. A set of the cyan circles with the error bars
 indicates the observed core positions at 22~GHz, while the other data set with
 the magenta circles represents the core positions at 43~GHz. We show all of the
 six epochs (2012 February 24, March 16, 21, April 5, 27 and May 17) obtained by
 the \textit{Mode-A} at each frequency. The error budget for each data point
 includes the following terms; the peak-to-noise ratio of the phase-referenced
 image of M~84, tropospheric residual, ionospheric residual and geometrical
 errors. We also plot a weighted mean position of the core over the six epochs as
 blue/red rectangles at 22/43~GHz, respectively. As the coordinate origin of this
 plot, the nominal phase-tracking center for M~87 $\alpha_{\rm J2000}={\rm
 12^{\rm h}30^{\rm m}49.\hspace{-.3em}^{\rm s}4233830}$ and $\delta_{\rm J2000}={\rm
 12^{\circ}23^{\prime}28.\hspace{-.3em}^{\prime\prime}043840}$ is used.}  \label{fig:}
\end{figure}

Although the position uncertainty estimated for each data point is not
sufficiently small (a level of 0.1--0.15~mas), our multi-epoch, collective data
set reveals a noticeable trend that the 22-GHz core of M~87 tends to be located
northeast of the 43-GHz core. At each frequency, we derived a weighted-mean
position of the M~87 core over the analyzed epochs to be $(x_{\rm 43}, y_{\rm 43})
= (-199\pm41, 25\pm41)~\mu$as (a red square in Figure~5) and $(x_{\rm 22}, y_{\rm
22}) = (-263\pm44, 120\pm45)~\mu$as (a blue square in Figure~5) relative to the
nominal phase-tracking center of M~87. These values yield a time-averaged
separation of the core between 22 and 43~GHz to be $(\Delta x_{\rm 22-43}$,
$\Delta y_{\rm 22-43}) = (64, 95)$~$\mu$as, corresponding to a projected distance
of (0.005, 0.008)~pc or (9, 13)~$R_{\rm s}$. This results in a position angle of
the core shift to be $326^{\circ}$, which is well within a wide opening angle
(between P.A.$\sim$$255^{\circ}$ and $\sim$$340^{\circ}$) observed at the M~87 jet
base with higher-angular-resolution studies~\citep{junor1999, hada2013}.

To be accurate, however, we should be cautious about the following additional
position uncertainties. As described in \citet{hada2011}, the structure of the
reference source M~84 is elongated in the north-south direction. This may produce
a core shift of M~84 itself in declination~\citep[although its amount should be
less than 10~$\mu$as between 22 and 43~GHz; see][]{hada2011}. Also, it is
empirically suggested that astrometric measurements with VERA tend to leave
slightly larger position errors in declination than in right ascension due to the
residuals of zenith tropospheric delay estimations~~\citep[e.g.,][]{honma2007,
hirota2007}. Thus, the actual position uncertainties of the M~87 core in
declination could be somewhat larger than those adopted in Figure~5.

Another concern is that the amount of the core shift obtained with VERA $(\Delta
x_{\rm 22-43}$, $\Delta y_{\rm 22-43})$ could be slightly larger than that of the
true core shift, since the ``radio core'' defined by VERA, a beam size of which is
typically $\sim$2--3 times larger than that of VLBA, would contain more amounts of
the optically thin jet emission beyond the $\tau_{\rm ssa}(\nu)\sim1$
surface. Such a blending of the optically thin jet should be more significant at
22~GHz as hinted by the result that the core spectra obtained by VERA tend to be
steeper than those of VLBA. Then, in order to estimate a potential (artificial)
increase of the core position shift due to the blending of the optically-thin jet,
we performed the following test; at 22~GHz, we convolved the M~87 structure by
many different beam sizes with diameters ranging from 0.6 to 2.4~mas in
incremental steps of 0.2~mas. Then, we plotted systematic changes of
brightness-peak position (i.e., effective core position at each resolution) as a
function of beam size. Using this plot, we found a progressive position shift of
the brightness peak (toward the downstream side) to be $(\sim6,\sim3)$~$\mu$as per
0.2~mas (in $(x,y)$ directions) for a range of beam size from 0.6 to 2.4~mas. This
implies that a core shift of $(\sim6,\sim3)\times\frac{\rm 0.6~mas}{\rm 0.2~mas}=
(\sim18,\sim9)~\mu$as can be artificially caused by the blending effect of the
unresolved optically thin jet within a beam size of 0.6~mas.

Therefore, we conclude that the true core shift of M~87 should be a level of
40--50~$\mu$as in the right ascension, while in declination 80--90~$\mu$as is
estimated with a slightly larger uncertainty. In terms of the component in right
ascension, a quite similar level of core shift is obtained in the previous
measurements with VLBA~\citep[$\sim$30~$\mu$as between 22 and
43~GHz;][]{hada2011}.

In principle, our multi-epoch astrometry allows us to investigate a time evolution
of the core position, which is particularly interesting to explore during a radio
flaring stage. However, the observed small position changes relative to the errors
make a reliable examination difficult among individual epochs. Conservatively, in
the this paper we can say that the core position is stable within a scale of
0.3~mas (corresponding to $\sim$40~$R_{\rm s}$ or $\sim$0.02~pc) during the radio
flare by taking into account the maximum extent of the core position scatters in
these data.

\section{Discussion}
\subsection{The location of the 2012 VHE activity}

The location and the origin of the $\gamma$-ray emission up to VHE from M~87 have
been under active debate in recent years. In the 2005 episode, the VHE flare was
accompanied by radio-to-X-ray flares from HST-1~\citep{harris2006} with the
ejections of superluminal radio features~\citep{cheung2007}, leading to a scenario
that the VHE emission originates in HST-1~\citep{stawarz2006, cheung2007,
harris2008, harris2009}. In contrast, contemporaneous flux increases at VHE, from
the X-ray core and from the 43-GHz core during the 2008 episode suggest that the
VHE flare originates in the jet base~\citep{acciari2009}. A coincident enhancement
of the X-ray core flux also occurred during the VHE flare in
2010~\citep{harris2011, abramowski2012} together with a possible flux density
increase from the radio core~\citep{hada2012}. These results also tend to favor
the scenario that the VHE emission originates in the jet base. However, the
detection of the new component emergence from HST-1 near the VHE event again
evoked the possibility that the HST-1 region is related to the active event in
2010~\citep{giroletti2012}.

During the period of the elevated VHE activity in 2012~\citep{beilicke2012}, we
detected a significant flux density increase from the radio core at both 43 and
22~GHz. Following the 2008 episode this is the second time where a VHE event
accompanied a remarkable radio flare from the core. Meanwhile, the radio
luminosity of the HST-1 region was continuously decreasing, and we did not find
any hints of the emergence of new components from HST-1 as seen in 2005 and
2010. These results strongly suggest that the VHE activity in 2012 is associated
with the core at the jet base, while HST-1 is an unlikely source. We should note
that these remarkable flares are very rare also in radio bands~\citep[see the
supporting material in][]{acciari2009}, so it is unlikely that an observed joint
radio/VHE correlation is a chance coincidence, while the low statistics of the LAT
light curves still do not allow conclusive results on the HE-VHE
connection. Moreover, we also emphasize the clear detection of a progressive
change of the radio core spectral index after the VHE high state using three
frequencies. This was not revealed in the previous radio flare during the 2008 VHE
event, where there was only single radio coverage at 43~GHz.

Currently a variety of mechanisms have been proposed for the VHE $\gamma$-ray
production from the jet base of M~87 based on both leptonic and hadronic
origins. In what follows, we will briefly discuss the compatibility of these
models with the VHE activity in 2012, specifically from the point of view of
high-resolution radio observations. On the basis of VLBI data, we can summarize
some of the key observational constraints as follows; first, the VERA 43-GHz
images indicate the size of the radio flaring region to be smaller than the
central spatial resolution of $\sim$0.4~mas (56~$R_{\rm s}$ or 0.03~pc), which is
consistent with a source size implied by the observed month-scale VHE variability
in 2012 ($R_{\rm var} \sim c\delta t_{\rm var}\sim 7.5\times 10^{16}\delta$ cm or
$42\delta~R_{\rm s}$ for $t_{\rm var}\sim30$~days, where $\delta$ is the Doppler
factor); second, M~87 has a very short BH-to-core distance~\citep[$\sim$a few tens
of $R_{\rm s}$;][]{hada2011} and the jet launch/collimation region is beginning to
be resolved at its base~\citep{junor1999, asada2012, hada2013}; third, while the
maximum of the radio flare is delayed, the onset of the radio brightening in 2012
occurs simultaneously with the VHE enhancement, indicating that the two emission
regions are not spatially separated. Therefore, any compatible scenarios with the
2012 VHE activity should satisfy these conditions obtained by the radio
observations.

Some of the existing models ascribe the VHE production to extremely compact
regions (BH magnetosphere; \citet{neronov2007, rieger2008, levinson2011}:
multi-blob in jet launch/formation region; \citet{lenain2008}: mini-jets in the
main jet; \citet{giannios2010}: or red-giant-stars/jet-base interactions;
\citep{barkov2010}). These models well explain the rapid (a few days) variability
observed in the previous VHE flares in 2005, 2008 and 2010. Also, these models are
able to avoid the internal $\gamma$ (VHE) -- $\gamma$ (IR) absorption problem near
the black hole since M~87 has only a weak IR nucleus~\citep{neronov2007,
brodatzki2011}. However, as far as we consider the case in 2012, the size of the
associated region expected from these models (typically of the order of $R_{\rm
s}$ but as small as $\sim$$0.01~R_{\rm s}$ in Lenain et al.~2008) seems to be
significantly smaller than that suggested by VLBI and the observed longer
timescale of the VHE variability. Note that a contemporaneous mm-VLBI observation
at 230~GHz during the 2012 event also suggests possible constraint on the
associated size to be a similar spatial scale ($\gtrsim$0.3~mas; Akiyama et al. in
prep.).

On a larger scale (but still unresolved with cm-VLBI), \citet{georganopoulos2005}
proposed a blazar-type, two-zone emission model where the VHE emission is produced
in the upstream, faster jet while the lower-energy emission originates in the
downstream, decelerating part of the jet. This model can reproduce the VHE data of
M~87 if the jet decelerates its bulk Lorentz factor from $\Gamma_{\rm j}\sim20$ to
$\Gamma_{\rm j}\sim5$ over a length of $\sim$0.1~pc at the jet base. However, the
previous VLBA monitoring studies as well as our VERA monitor often show
sub-relativistic motions and do not support such a very fast speed near the jet
base~\citep{reid1989, dodson2006, ly2007, kovalev2007}, although one cannot
completely rule out faster motions in the jet base region that cannot be resolved
with current cm-VLBI. The application of this model for the jet base of M~87
(i.e., jet launch region within $\lesssim100~R_{\rm s}$ from the black hole) seems
to be also incompatible with the current theoretical paradigm on the jet
production, where the launched jet is expected to start acceleration on this
scale~\citep[e.g.,][]{mckinney2006, komissarov2007}.

Another blazar-type scenario in the sub-parsec jet is proposed by
\citet{tavecchio2008}, where the radio-to-GeV emission of M~87 originates in the
interior, spine part of the jet while the VHE emission is produced in the
surrounding layer. However in their steady state model, whether the model can
explain the observed simultaneous radio/VHE correlation or not has not been well
investigated yet because the emission regions associated with radio and VHE are
spatially separated from each other.

Similarly, the existing hadronic-based VHE emission
models~\citep[e.g.,][]{reimer2004, barkov2010, reynoso2011} do not necessarily
expect obvious radio/VHE correlations because the radio emission is usually
ascribed to be the synchrotron emission from the electrons that are not
responsible for the VHE production. To test the viability of these (both
multi-zone and multi-particle) models, more detailed considerations including
their time-dependent behaviors through radio to VHE $\gamma$-ray would be
necessary.

Interestingly, \citet{abdo2009} showed that a broadband SED of M~87 through radio
to VHE (at a relatively moderate state) can be reasonably reproduced by a simple,
homogeneous one-zone synchrotron self-Compton (SSC) jet model using a roughly
comparable source size ($\lesssim$0.1~mas) to the one suggested in the 2012
case. In principle, one can also accept coincident radio/VHE correlations in this
context. One-zone SSC models also explain the broadband SEDs in the other known
$\gamma$-ray detected FR-I radio galaxies 3C\,84/NGC\,1275~\citep[up to
TeV;][]{magic2013} and Centaurus~A~\citep[up to GeV;][]{chiaberge2001, abdo2010},
implying these sources to be misaligned counterparts of BL Lac objects. However,
such single-zone approaches to misaligned objects derive systematically smaller
bulk Lorentz factors ($\Gamma_{\rm j}<5$) than those generally required in BL Lacs
($\Gamma_{\rm j}>10$), which may pose a conflict with the radio-loud AGN
unification paradigm~\citep{urry1995}. This caveat, along with limb-brightened jet
structures observed in some radio galaxies~\citep[e.g.,][]{junor1999, nagai2014},
indeed led to the preferable use of more sophisticated models with multiple jet
parts at different velocities~\citep[e.g.,][]{georganopoulos2005, tavecchio2008}
or with multiple substructures~\citep[e.g.,][]{lenain2008, giannios2010}. For
M~87, such multi-zone models also should be viable, but more extended theoretical
consideration would be needed (e.g., ways to produce the joint radio/VHE
correlation).

\subsection{The nature of the radio flare}
Our multi-frequency radio monitoring revealed a frequency-dependent property of
the radio light curves for the M~87 core; as frequency decreases, the luminosity
slowly increases with a smaller amplitude enhancement, resulting in a larger time
delay/lag until reaching its peak. While this type of behavior is often seen in
blazars~\citep[e.g.,][]{orienti2013}, the present study is the first clear
confirmation for the M~87 jet. Such a behavior is often explained by the creation
of a plasma condensation (presumably induced by the VHE event at the jet base),
which subsequently expands and propagates down the jet under the effect of
SSA~\citep{marscher1985, valtaoja1992}. In this idea, the stronger SSA opacity at
the jet base causes a delayed brightening of the radio flux density at lower
frequencies. The light curve at each frequency reaches its maximum when the
newborn component passes through the $\tau_{\rm ssa}(\nu)\sim 1$ surface (i.e.,
the radio core at the corresponding frequency). Afterward, the light curve enters
a decay phase due to the adiabatic expansion loss (plus additional cooling
effects, if any). A self-absorbed synchrotron response of radio core light curve
in M~87 is also suggested for the 2008 episode \citep{acciari2009}, where the
delayed increase of the 43-GHz core flux relative to the VHE peak was reasonably
reproduced by a phenomenological modeling of a self-absorbed plasma at the jet
base.

On the other hand, we did not see any flux enhancement from the 5~GHz core of
M~87. This implies that the newborn component was short-lived and incorporated
into the uniform (underlying) jet before reaching the $\tau_{\rm ssa}(5~{\rm
GHz})\sim 1$ surface, resulting in the absence of any explicit emergences of knots
or notable features from the core during this period. We note that the inner jet
region of M~87 shows a relatively smooth brightness distribution, in agreement
with low power (FR Is and BL Lacs) jet properties, while jets in high power
sources (FR IIs and quasars) show clear evidence of complex
substructures~\citep{giovannini2001}.

One of the important results of our observations is that we independently obtained
both the core shift and the time-lag of the multi-frequency flare
simultaneously. As examined by \citet{kudryavtseva2011}, the time-lag of the radio
flare is directly related to the amount of core shift divided by the speed of the
propagating shock or component. In turn, this means that we can now estimate the
speed of the propagating component in the following way using the core shift and
the time-lag;
\begin{eqnarray*}
\beta_{\rm app, 43 \rightarrow 22} &=& \frac{\Delta r_{\rm proj, 43-22}}{c \Delta
 t_{\rm 43-22}} \\
 &=& 0.97 \left(\frac{\Delta\theta_{\rm 43-22}}{\rm 0.1~mas}\right) \left(\frac{\Delta t_{\rm
      43-22}}{\rm 10~days}\right)^{-1} \left(\frac{D_{\rm source}}{\rm 16.7~Mpc}\right)
\end{eqnarray*}
where $\beta_{\rm app, 43 \rightarrow 22}$, $\Delta r_{\rm proj, 43-22}$,
$\Delta\theta_{\rm 43-22}$, $\Delta t_{\rm 43-22}$ and $D_{\rm source}$ represent
a proper (projected) motion of the propagating component from the 43-GHz core to
the 22-GHz one, a projected separation of 43/22~GHz cores (in physical unit/in
angular unit), a time-lag of the light curve peaks between 43 and 22~GHz, and a
source distance, respectively. For $\Delta\theta_{\rm 43-22}$, our VERA astrometry
obtained a level of $\Delta\theta_{\rm 43-22}$$\sim$0.04--0.09~mas during the
radio flare. Regarding $\Delta t_{\rm 43-22}$, although the obtained light curves
are not adequate to specify the exact date of the peak at both 22 and 43~GHz, we
can still reasonably constrain a likely range of the peak date to be between
MJD~56044 and MJD~56064 at 43~GHz, and between MJD~56099 and MJD~56168 at 22~GHz,
respectively. This results in the maximum allowed range of $\Delta t_{\rm 43-22}$
to be $35~{\rm days} \lesssim \Delta t_{\rm 43-22} \lesssim 124~{\rm
days}$. Therefore, we finally derive a likely range of $\beta_{\rm app, 43
\rightarrow 22}$ to be $\sim$0.04$c$--0.22$c$. This indicates that the newborn
component propagating through the cores between 43 and 22~GHz is not highly
relativistic. Notably, many previous kinematic observations of M~87 obtained a
similar speed \citep[$\lesssim 0.05~c$;][]{kovalev2007} or a slightly higher
value~\citep[$0.25-0.4~c$; ][]{ly2007} in the jet a few mas (subparsec) downstream
of the core.  A value around $\sim$0.4$~c$ (relative to the core) at a few mas
beyond the core was also measured in our VERA data~\citep{hada2013b}, suggesting a
possible acceleration process in the jet flow on this scale. On the other hand,
the derived sub-luminal speed of the newborn component is significantly smaller
than the super-luminal features appeared from the core during the previous VHE
event in 2008~\citep[$\sim$1.1$c$;][]{acciari2009}, where the peak VHE flux is
$>$5 times higher than that in 2012. If we assume that propagating shocks or
component motions traced by radio observations reflect the bulk velocity flow,
this may suggest that the stronger VHE activity is associated with the (episodic)
production of the higher Lorentz factor jet. We note that the M~87 jet is
misaligned~\citep[$i \sim 15^{\circ}-25^{\circ}$;][]{acciari2009}, but such a
positive correlation between the jet speed and the observed VHE flux is plausible 
because the jet emission is still ``beamed'' toward us ($\delta \sim$a few) for
such mildly relativistic speeds ($\Gamma \lesssim$2) in the suggested $i$ range.

We note that the above discussion applies to a thin-shock case where the thickness
(i.e., the size along the jet) of the propagating feature is smaller than the
(de-projected) separation between the 43 and 22-GHz cores in order to allow a
non-zero time delay ($\Delta t_{\rm 43-22}$) in their light curve peaks. Also,
here we considered only a time-averaged value for the core shift for
simplicity. However, the amount of core shift can be variable especially during
the radio flaring event~\citep{kovalev2008}, because the creation of such features
may locally change the particle number density, the magnetic field strength and
thus the SSA optical depth at the jet base. The position accuracies in the present
study are still not enough to explore the core-shift variations among the
individual epochs, but this effect should be of interest to test with a more
suitable astrometry technique~(e.g., Korean VLBI Network (KVN); Lee et al. 2014:
or KVN+VERA joint array; Sawada-Satoh 2013). This will enable us to bring further
insights into the nature of the flaring core as well as the $\gamma$-ray
production in the formation and collimation region of the M~87 jet.

\section{Summary}
We reported our intensive, high-resolution radio monitoring observations of the
M~87 jet with VERA at 22/43~GHz and EVN at 5~GHz during the elevated VHE state in
early 2012 jointly with the contemporaneous MeV/GeV light curves obtained by
$Fermi$-LAT. We summarize our main results as follows;

\begin{enumerate}
\item We detected a strong increase of the radio flux density from the jet base
      (radio core) at both 22 and 43~GHz coincident with the elevated VHE
      activity. Meanwhile, we confirmed that the HST-1 region remained quiescent
      in terms of its flux density and mas-scale structure. These results strongly
      suggest that the VHE event in 2012 originates in the jet base within the
      central resolution element of $0.4~{\rm mas}=0.03~{\rm pc}=56~R_{\rm s}$
      near the supermassive black hole, while HST-1 is unlikely.

\item We discovered a clear frequency-dependent evolution of the radio core light
      curves at 43, 22 and 5~GHz. The radio flux increased more rapidly at 43~GHz
      with a stronger amplitude, while the observed light curve at 5~GHz remained
      stable. Taking advantage of the dual-beam astrometry technique with VERA, we
      also detected a core shift between 22 and 43~GHz. These results suggest that
      a new radio-emitting component was created near the black hole during the
      VHE event, and propagating through the opaque cores at these
      frequencies. Since we did not see any flux/structural variations for the
      5-GHz core, this implies that the newborn component was quenched before
      reaching the 5-GHz core. This confirms that jets in low power radio sources
      generally show a uniform brightness distribution because knotty features are
      short-lived and disappear very soon.

\item In terms of the MeV/GeV regime, the LAT light curves were stable during the
      analysed period before the VHE event, and we did not find any significant
      flux enhancement during the period of the VHE activity. Instead, we detected
      a factor of $\sim2$ decrease in the 0.1-100 GeV flux in 6-month intervals
      before and after the March 2012 VHE event, suggesting a state change also in
      the HE regime.

\item By combining the independent measurements of the time-lag in the radio light
      curves and the core shift between 22 and 43~GHz, we deduced a likely
      (apparent) speed of the newborn component within the location of the 22-GHz
      radio core (the physical scale less than $0.1~{\rm pc}$). We derived a
      sub-relativistic speed of less than $\sim$0.2$c$, being consistent with the
      values reported in previous kinematics measurements in the subparsec region
      of this jet. On the other hand, the derived speed for the newborn component
      is significantly slower than that of the component ($\sim$1.1$c$) that
      appeared during the VHE event in 2008. Considering the 2008 VHE activity to
      be more powerful than the case in 2012, this implies that the stronger VHE
      activity can be associated with the production of the higher Lorentz factor
      jet.
\end{enumerate}

\bigskip
\bigskip

We thank the anonymous referee for his/her review and suggestions for improving
the paper. We are grateful to E.~Torresi for reading and helpful comments on the
manuscript. The VERA is operated by Mizusawa VLBI Observatory, a branch of
National Astronomical Observatory of Japan. K.H. thanks K. M. Shibata, T. Jike and
all of the rest of the staff who helped with operations of the VERA observations
presented in this paper. K.H. is supported by the Japan Society for the Promotion
of Science (JSPS) Research Fellowship Program for Young Scientists. Part of this
work was done with the contribution of the Italian Ministry of Foreign Affairs and
University and Research for the collaboration project between Italy and
Japan. This work was partially supported by KAKENHI (24340042, 24540240, 24540242,
25120007 and 26800109). e-VLBI research infrastructure in Europe is supported by
the European Union's Seventh Framework Programme (FP7/2007-2013) under grant
agreement no. RI-261525 NEXPReS. The European VLBI Network is a joint facility of
European, Chinese, South African and other radio astronomy institutes funded by
their national research councils. The Submillimeter Array is a joint project
between the Smithsonian Astrophysical Observatory and the Academia Sinica
Institute of Astronomy and Astrophysics and is funded by the Smithsonian
Institution and the Academia Sinica.

The \textit{Fermi} LAT Collaboration acknowledges generous ongoing support from a
number of agencies and institutes that have supported both the development and the
operation of the LAT as well as scientific data analysis. These include the
National Aeronautics and Space Administration and the Department of Energy in the
United States, the Commissariat \`a l'Energie Atomique and the Centre National de
la Recherche Scientifique Institut National de Physique Nucl\'eaire et de Physique
des Particules in France, the Agenzia Spaziale Italiana and the Istituto Nazionale
di Fisica Nucleare in Italy, the Ministry of Education, Culture, Sports, Science
and Technology (MEXT), High Energy Accelerator Research Organization (KEK) and
Japan Aerospace Exploration Agency (JAXA) in Japan, and the K.~A.~Wallenberg
Foundation, the Swedish Research Council and the Swedish National Space Board in
Sweden.

Additional support for science analysis during the operations phase is gratefully
acknowledged from the Istituto Nazionale di Astrofisica in Italy and the Centre
National d'\'Etudes Spatiales in France.

\end{document}